\documentclass[aps,prb,twocolumn,showpacs,10pt]{revtex4-1}

\usepackage{amssymb,amsmath,bm,graphicx,times,tabularx}

\begin{document}

\title{The effect of quenched bond disorder on first-order phase transitions}

\author{Arash Bellafard}
\author{Sudip Chakravarty}
\affiliation{Department of Physics and Astronomy, University of California, Los Angeles, CA 90095, USA}

\author{Matthias Troyer}
\affiliation {Theoretische Physik, ETH Zurich, CH-8093 Zurich, Switzerland}

\author{Helmut G.~Katzgraber}
\affiliation {Department of Physics and Astronomy, Texas A\&M University,
College Station, Texas 77843-4242, USA}
\affiliation {Materials Science and Engineering Program, Texas A\&M
University, College Station, Texas 77843, USA}
\affiliation{Santa Fe Institute, 1399 Hyde Park Road, Santa Fe, New Mexico 87501 USA}
\date{\today}

\begin{abstract}
We investigate the effect of quenched bond disorder on the two-dimensional three-color Ashkin-Teller model, which undergoes a first-order phase transition in the absence of  impurities. This is one of the simplest and striking models in which quantitative numerical simulations can be carried out to investigate emergent criticality due to disorder rounding of first-order transition. Utilizing  extensive cluster Monte Carlo simulations on large lattice sizes of up to $128 \times 128$ spins, each of which  is represented  by three colors taking values $\pm 1$, we show that the rounding of the first-order phase transition is an emergent criticality. We further calculate the correlation length critical exponent, $\nu$, and the magnetization critical exponent, $\beta$, from finite size scaling analysis.  We find that the critical exponents, $\nu$ and $\beta$, change as the strength of disorder or the four-spin coupling varies, and we show that the  critical exponents appear not to be in the Ising universality class. We know of no analytical approaches that can explain our non-perturbative results. However our results  should inspire further work on this important problem, either numerical or analytical.
\end{abstract}


\maketitle

\section{Introduction}
Disorder is an inevitable part of any condensed matter system and therefore its study has always been of great importance.
The effect of randomness on the transition temperature (random $T_{c}$) and randomness coupled to the site variables  on  continuous phase transitions have been extensively studied for a long time, and they appear to be  well understood.

For random $T_{c}$ model defined above, the Harris criterion~\cite{harrisJPC1974} predicts conditions under which disorder can change  the universality class of the pure system.
He showed that  if the specific heat  exponent, $\alpha$, of a pure system is positive, i.e., if the product of the correlation length critical exponent, $\nu$, and the dimension of the system, $D$, is less than $2$ ($D\nu < 2)$, the effect of impurity is relevant: the pure system's fixed point is unstable, the system's critical exponents change, and the disordered system does not remain in the universality class of the pure system. On the other hand, if the specific heat critical exponent, $\alpha$, of a pure system is negative, i.e., if $D\nu > 2$, the effect of impurity is irrelevant: the pure system's fixed point is stable, critical exponents do not  change, and the disordered system remains in the universality class of the pure system. If $\alpha = 0$, the situation is marginal. 

The  problem of random field coupled linearly to the order parameter is  different, however. Using the domain-wall argument, Imry and Ma~\cite{imryPRL1979} showed that, for a $D$ dimensional system with discrete order parameter, the stability condition of  order  requires  $D/2 \le D-1$. For the continuous case the corresponding stability condition is   $D/2 \le D-2$. The marginal cases are treated in Refs.~\onlinecite{aizenmanPRL1989,aizenmanCMP1990}. The increase of critical dimensionality is verified for the random field Ising model in Ref.~\onlinecite{aharonyPRB1978}; for a review of this model, see Ref.~\onlinecite{natterman1997}.

First-order transitions are  ubiquitous in both classical and quantum systems, because they do not require any fine tuning of the coupling constant. However, in contrast to the effect of disorder on continuous transitions much less is known about its effect on first-order  transitions.
Rounding of first-order phase transition due to  quenched impurities that couple to energy-like variables has been studied in the past, yet the results are still not fully elucidated.
In an early study, \citeauthor*{imryPRB1979}~\cite{imryPRB1979} made use of Imry-Ma~\cite{imryPRL1979} domain-wall argument and showed that the presence of quenched bond randomness may produce rounding of a first-order phase transition. This happens because bond randomness couples to the local energy density of the system the same way that the random field couples to the local magnetization.
Using a renormalization-group calculation, \citeauthor*{huiPRL1989}~\cite{huiPRL1989,berkerPA1993} confirmed this idea  and showed that, for the $q$-states Potts model, the bond randomness turns a first-order phase transition into a second order transition. In another work, \citeauthor*{aizenmanPRL1989}~\cite{aizenmanPRL1989,aizenmanCMP1990} rigorously proved the elimination of discontinuity in the density of the variable conjugate to the fluctuating order parameter. Specifically, they showed the absence of the latent heat for the $q$-state random bond Potts model.
The effect of quenched bond randomness  on quantum systems that undergo a  first-order  phase transition in the pure case has also been touched upon, but without any firm conclusions involving the nature of the criticality and critical exponents.~\cite{goswamiPRL2008,greenblattPRL2009,greenblattPA2010,aizenmanJMP2012,hrahshehPRB2012}

We list further studies~\cite{chenPRL1992,wisemanPRE1995,cardyPRL1997,cardyJPA1996,pujolEPL1996} in two-dimensions ($2D$). The questions to be answered are whether or not the rounding is an emergence of criticality, i.e., does the correlation length diverge? If so, what are the exponents  and what are  the   universality classes, if any? The $q=8$ state random bond Potts model,~\cite{chenPRL1992} which has first order transition in the pure system,  hinted on the consistency with the  universality class of the pure $2D$ Ising model, which is known to have $\nu=1$ and $\beta=1/8$. 
However, in a later study of the critical behavior of the random bond Potts model it was found~\cite{cardyPRL1997}  that although the correlation length critical exponent $\nu$ is numerically close to unity (Ising), the magnetic exponent $\beta/\nu$ is far from the value $1/8$ and varies continuously with $q$, and therefore the disordered system cannot be in the universality class of the pure Ising model.~\cite{jacobsenNPB1998,olsonPRB1999,piccoPRL1997} Similar behavior was also observed in the study by \citeauthor*{chatelainPRL1998}.~\cite{chatelainPRL1998} 

As to  one~\cite{cardyJPA1996}  and two-loop~\cite{pujolEPL1996} {\em perturbative  renormalization group calculations},  $N$-color  AT model  hinted at the Ising universality class, contrary to our present work, as well as our recent work in  smaller lattices, $32\times 32$.~\cite{bellafardPRL2012} The validity of such perturbative calculations can of course be  doubted, as the renormalization group trajectories flow to strong coupling before curling back to the pure $N$-decoupled Ising fixed points.

%
Our previous work~\cite{bellafardPRL2012} could also be doubted as to whether or not  the observed behavior is an artifact of finite size effects.
To address this issue, it is important to do the calculations on larger system sizes. In this paper, we provide more precise results obtained from an extensive cluster Monte Carlo calculation on lattice sizes of up to 16 times larger in area ($128 \times 128$). Furthermore, we calculate the value of $\nu$ by finite-size scaling of two different quantities: (1) the logarithmic derivative of the magnetization and (2) the magnetic cumulant.~\cite{binderPRL1981} It is striking that the obtained value of $\nu$ still violates the  lower bound, $2/D$, and that the values of $\nu$ and $\beta$ change as the strength of disorder or the four-spin coupling varies.

The outline of this paper is as follows: in Section \ref{sec:the_model}, we introduce the $N$-color AT model and the binary  bond disorder. In Section \ref{sec:computational_method}, we explain the cluster MC method that we utilize for the analysis of the three color AT model. Thereafter, we show our results and provide a discussion of our findings in Sections \ref{sec:results} and \ref{sec:conclusion}, respectively.

\section{The Model}\label{sec:the_model}
To provide a brief background, a model that can shed light on {\em both} the classical and the quantum versions of the $N$-color AT model is the massive Gross-Neveu model in (1+1) dimensions~\cite{Gross:1974} with random mass.~\cite{Dotsenko:1985} Consider first the pure model. The fundamental dynamical variables are Dirac fields $\psi^{\alpha}$,  $\alpha=1,\ldots N$. In two dimensions the Dirac fields have only two components and the Dirac matrices are $2\times 2$ matrices. In the usual representation, $\gamma^{0}=\sigma_{z}$, $\gamma^{1}=i\sigma_{x}$, and $\gamma^{5}=\gamma^{0}\gamma^{1}=\sigma_{x}$, where the $\sigma$'s are the Pauli matrices. In all other respects, the conventions are the same as in four dimensions. The Lagrangian is
\begin{equation}
{\cal L}=\overline{\psi}^{\alpha}i\partial_{\mu}\gamma^{\mu}\psi^{\alpha}-m_{0}\overline{\psi}^{\alpha}\psi^{\alpha}+g_{0}(\overline{\psi}^{\alpha}\psi^{\alpha})^{2}.
\end{equation}
One can see that the corresponding Euclidean action is equivalent to the continuum limit of a two-dimensional model with $N$-colors, $\{s^{\alpha}_{i}=\pm 1: (\alpha=1, \ldots, N)\}$, with four-spin interactions that reflect the coupling between the energy densities. The Hamiltonian is
\begin{equation}
H = -\sum_{\langle i,j\rangle }^{L} J \sum_{\alpha}^{N} s_{i}^{(\alpha)}s_{j}^{(\alpha)} - g\sum_{\langle i,j\rangle}^{L} \sum_{\alpha\ne \beta}^{N} s_i^{(\alpha)} s_j^{(\alpha)} s_i^{(\beta)} s_j^{(\beta)}.
\label{eq:hamiltonian}
\end{equation}
Here $\langle i,j\rangle $ corresponds to nearest-neighbor lattice sites.  For $N=2$, the model is the same as the ``standard'' Ashkin-Teller model~\cite{ashkinPR1943,fanPLA1972} with a line of fixed points labeled by $g$.

For $g=0$, it represents $N$ decoupled Ising models. It is in the vicinity of this Ising transition, $T_{c}$, that the continuum limit was constructed, so that $m_{0}\sim |1-T/T_{c}|$.
If $g < 0$ and $N \ge 3$, the four-spin coupling term is marginally irrelevant and the Hamiltonian, regardless of disorder, exhibits a continuous phase transition (CPT) and there exists a line of fixed points along which the critical exponents vary continuously. On the other hand, if $g > 0$, with ferromagnetic coupling, $J > 0$, and $N \ge 3$, the four-spin coupling term is marginally relevant and the pure Hamiltonian undergoes a first-order phase transition. This has been shown by a large-$N$ analysis,~\cite{fradkinPRL1984} mean-field theory,~\cite{grestPRB1981} perturbative RG analysis,~\cite{grestPRB1981} numerical simulations,~\cite{grestPRB1981} and general arguments.~\cite{shankarPRL1985} For the rest of this paper, we will focus only on $g > 0$, the ferromagnetic three-color case.

Now imagine that we replace $J$ by $J\to J+\delta J_{ij}$, where the random variable $\delta J_{ij}$ satisfies the impurity average $\overline{\delta J_{ij}}=0$ and $\overline{(\delta J_{ij})^{2}}=W_{0}^{2}$. The bond disorder maps onto  mass disorder of the fermion action of the Gross-Neveu model, but this corresponds to isotropic white noise disorder in both space and imaginary
time directions.

The generalization to the corresponding quantum problem will involve only spatial disorder in the quantum model, which translates into a highly anisotropic disorder in the $N$-color classical version. The exchange constants in the vertical direction (imaginary time) are identical to each other within each column but different from column to column, the same as in the McCoy-Wu problem.~\cite{McCoy:1968,Fisher:1995}

However, in the present paper, we do not study the quantum problem, but only the classical problem:
\begin{equation}
{\cal H} = -\sum_{\langle i,j\rangle }J_{ij}\sum_{\alpha}s_{i}^{(\alpha)}s_{j}^{(\alpha)}-g\sum_{\langle i,j\rangle}\sum_{\alpha\ne \beta}s_i^{(\alpha)} s_j^{(\alpha)} s_i^{(\beta)} s_j^{(\beta)}.
\label{eq:ham}
\end{equation}
The corresponding quantum problem is more difficult to simulate in large systems and brings in the additional questions about activated scaling, which was recently explored in $(1+1)$-dimensions.~\cite{bellafardARX2014} The classical problem, however, is sufficient to address the most basic questions about emergent criticality in a disorder rounded first-order transition.

%

Because the critical behavior of the system should be independent of the choice of disorder, the easiest disorder one can introduce to this system is the binary bond disorder:
\begin{equation}
	p[J_{ij}] = \begin{cases} J+\Delta/2, \text{~with probability 1/2}\\ J-\Delta/2, \text{~with probability 1/2} \end{cases}
\end{equation}
Moreover, with binary disorder distribution one needs to average over fewer disorder configurations in order to achieve reliable simulation results. We also do not introduce disorder in the inter-color four-spin coupling $g$.

\section{Computational Method}\label{sec:computational_method}
We use the cluster MC method for our calculations. Since single flip MC suffers from slowing down at a phase transition due to the increase of fluctuation, correlation time, and, for  CPT,  the divergence of correlation length close to the critical point, we resort to cluster MC which handles these problems much better. Fixing a single color, for instance $1$, in the bond-disordered version of the above Hamiltonian, Eq.~\eqref{eq:ham}, we can write
\begin{align}
	\mathcal H
	&= \mathcal H_{\bar 1} - \sum_{\langle i,j \rangle} \left(J_{ij} + g \sum_{\alpha\neq 1} s_i^\alpha s_j^\alpha\right) s_i^1 s_j^1
	\label{eq:modHamiltonian}
\end{align}
where the first term, $\mathcal H_{\bar 1}$, does not contain the color 1. The second term of the Hamiltonian can be regarded as the Hamiltonian of the Ising model with coupling constant $J_{ij} + g \sum_{\alpha\neq 1} s_i^\alpha s_j^\alpha$.
Therefore, we can implement any sort of cluster MC algorithm suited for the Ising model.
We choose the cluster MC suggested by Niedermayer~\cite{niedermayerPRL1988} which is a generalization of Swendsen-Wang cluster MC.~\cite{swendsenPRL1987}

For a randomly chosen color, we randomly choose a lattice site that is hosting a spin. The site is now the single member of the cluster. We let the cluster grow by adding all neighbors of the selected site with probability
\begin{equation}
	P_{\rm add}(E_n) = 1 - e^{E_n-E_{\rm max}}
\end{equation}
where $E_n$ is the energy for bond $n$ given by the latter part of Eq.~\eqref{eq:modHamiltonian} and $E_{\rm max} = \max(J_{i,j}) + (N-1) g = J + \Delta/2 + (N-1)g$ is the upper bound of the bond energy. Ergodicity is trivially satisfied since there will always be a nonzero probability for which the cluster consists of only one single site. The cluster growth process is repeated until no neighboring sites can be added to the cluster. Then, the entire spins of the cluster are flipped.

The simulations have been done at $J$'s close to the expected  critical point  $J_c$ for all sizes. The ``thermalization time'' was estimated using logarithmic binning method, i.e., by comparing the average values of each observable over $2^n$ MC steps and requiring that the last three averages be within each others error bars. As a result, the systems were updated more than ten million times before the equilibrium was reached. We have averaged each observable over $2000$ to $20000$ disorder configurations and for each disorder configuration, we have performed $10000$ thermal averages. The number of disorder configurations is given in the caption of the plots. The observables' error bars are calculated using the Jacknife procedure.~\cite{wuAOS1986,Young:2012}

\section{Results}\label{sec:results}
We make use of the lowest-order energy cumulant,~\cite{challaPRB1986,binderPRB1984} given by
\begin{equation}\label{eq:energy_cumulant}
	V_E = 1 - \frac{[\langle E^4 \rangle]}{3[\langle E^2 \rangle]^2},
\end{equation}
to find the order of the phase transition. Here $E$ is the energy density per spin, per unit area. The angular brackets $\langle\dots\rangle$ denote the usual thermal MC average, whereas the square brackets $[\dots]$ denote the quenched average over configurations with different $\{J_{ij}\}$.

Away from the phase transition point, $J_c$, the probability distribution of the energy is a $\delta$-function in the thermodynamic limit, therefore~\cite{challaPRB1986}
\begin{equation}\label{eq:ve_trivial_limit}
	V_E \to \frac{2}{3}, \quad L \to \infty, \quad J \neq J_c \quad \text{fixed}.
\end{equation}
At $J = J_c$, $V_E$ behaves differently for first- and second-order phase transitions.
If the phase transition is continuous, the probability distribution of the energy is a $\delta$-function in the thermodynamic limit,~\cite{challaPRB1986} and
\begin{equation}
	V_E \to \frac{2}{3}, \quad L \to \infty, \quad \text{at}~ J = J_c.
\end{equation}
If the phase transition is first-order, the probability distribution of the energy is described by two Gaussians of equal weight.~\cite{challaPRB1986} Therefore,
\begin{equation}
	V_E|_\text{min} \to \text{const.} \neq \frac{2}{3}, \quad L \to \infty, \quad J = J_c(L).
\end{equation}

\begin{figure}[tbp]
	\begin{center}
		\includegraphics[width=\linewidth]{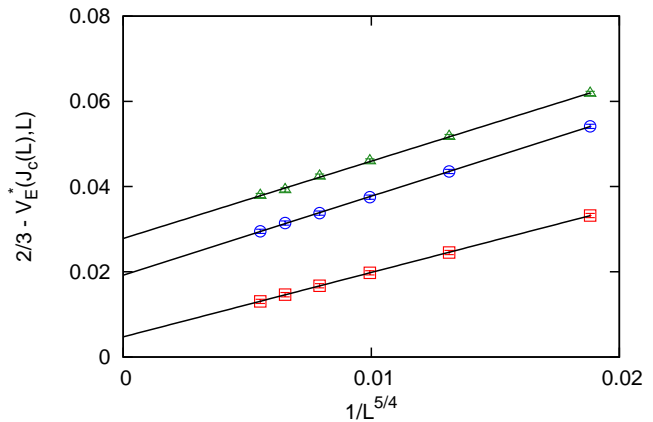}
	\end{center}
	\caption{(color online). $[2/3 - V_E^*(J_c(L))]$ versus $1/L^{\rm const}$ for $N = 3$. From top to bottom, $(g,\Delta) = (0.12,0), (0.1,0),$ and $(0.08,0)$ with the ordinate intercepts at $0.0278(2)$, $0.0192(1)$, and $0.0047(1)$. The system sizes are $L = 24, 32, 40, 48, 56,$ and $64$. The measured values are averaged over $20000$ configurations. The fitted lines intersect the ordinate at a finite non-zero values indicating a first-order phase transition.}
	\label{fig:gd00veStar}
\end{figure}

We now try to find the order of the phase transition for the pure system, i.e., when $\Delta = 0$. We fix $g$ and calculate $V_E$ for the system sizes $L = 24, 32, \dots, 64$. We plot $2/3 - V_E^*[J_c(L),L]$ as a function of $L^{5/4}$. Fig.~\ref{fig:gd00veStar} shows the results for $g = 0.08, 0.10,$ and $0.12$. We choose the power of $L$ to be $5/4$ because it gives us the best linear fit to our data. All fitted lines intersect the ordinate at a non-zero finite value which indicates that the pure system undergoes a first-order phase transition in agreement with earlier works.~\cite{fradkinPRL1984,grestPRB1981,bellafardPRL2012}

\begin{figure}[tbp]
	\begin{center}
		\includegraphics[width=\linewidth]{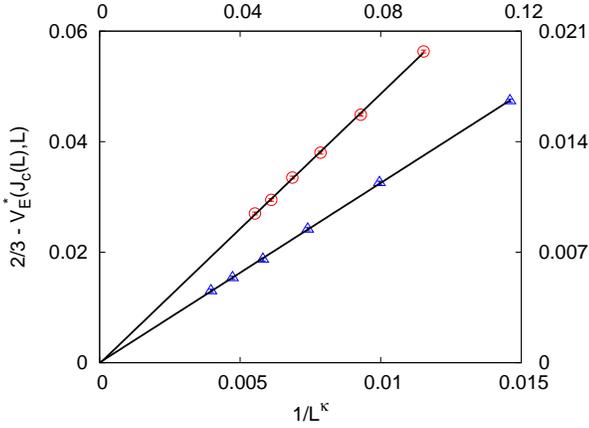}
	\end{center}
	\caption{(color online). A plot of $[2/3 - V_E^*(J_c(L))]$ versus $1/L^{\kappa}$. The left and top axes correspond to the round symbols with the parameter set $(g,\Delta) = (0.1,0.2)$ and $\kappa = 3/4$. The fitted line intersects the ordinate at $0.0002(4)$. The right and bottom axes correspond to the triangular symbols with $(g,\Delta) = (0.05,0.2)$ and $\kappa = 4/3$. The fitted line intersects the ordinate at $0.00001(5)$. The system sizes are $L = 24, 32, 40, 48, 56,$ and $64$. The measured values are averaged over $15000$ configurations. The error bars are smaller than the symbol sizes. The fitted lines go through the origin indicating a continuous phase transition.}
	\label{fig:gd20veStar}
\end{figure}

Now, we turn our attention to the quenched bond disordered system, i.e., when $\Delta \neq 0$. We calculate the energy cumulant, $V_E$, of the system for different values of $g$ and $\Delta$. Again, we plot the depth of the energy cumulant, $2/3 - V_E^*[J_c(L),L]$, as a function of some inverse power of the system size, $L^{-\kappa}$, as shown in Fig.~\ref{fig:gd20veStar}. This time, we observe that the fitted lines go through zero. We conclude that the phase transition in the presence of the quenched bond disorder is continuous. The inverse powers of the system sizes are different for different parameter sets ($\kappa = 3/4$ for $(g,\Delta) = (0.1,0.2)$ and $\kappa = 4/3$ for $(g,\Delta) = (0.05,0.2)$) and we choose them such that we get the best fit to our data.

\begin{figure}[tbp]
	\begin{center}
		\includegraphics[width=\linewidth]{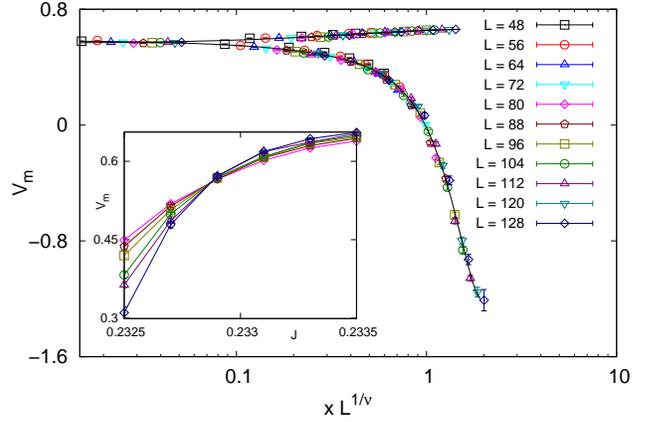}
	\end{center}
	\caption{(color online). Data collapse of the magnetic cumulant for the parameter set $(g,\Delta) = (0.1,0.2)$ with $N = 3, J_c = 0.23287$ and $\nu = 0.81$ for $L = 48, 56, \dots, 128$. The upper (lower) branch corresponds to the ordered (disordered) phase. The inset shows the magnetic cumulant $V_m$ versus $J$. The measured values for system sizes of up to $L = 112$ are averaged over $5000$ configurations. For $L = 120$ and $128$ the measured values are averaged over $2000$ configurations. The intersection point indicates that the critical point is at $J_c = 0.23287(3)$.}
	\label{fig:g10d20vm}
\end{figure}

\begin{figure}[tbp]
	\begin{center}
		\includegraphics[width=\linewidth]{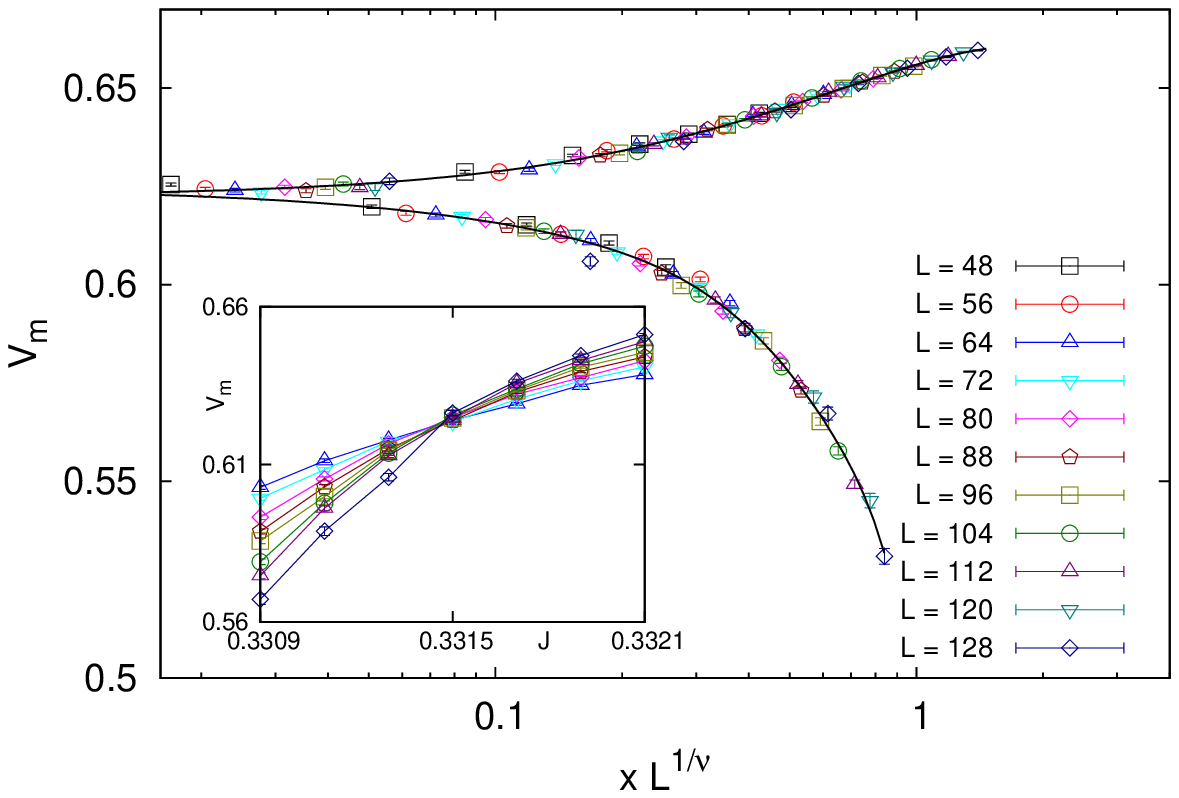}
	\end{center}
	\caption{(color online). Data collapse of the magnetic cumulant for the parameter set $(g,\Delta) = (0.05,0.2)$ with $N = 3, J_c = 0.33145$ and $\nu = 0.82$ for $L = 48, 56, \dots, 128$. The upper (lower) branch corresponds to the ordered (disordered) phase. The inset shows the magnetic cumulant $V_m$ versus $J$. The measured values for system sizes of up to $L = 112$ are averaged over $5000$ configurations. For $L = 120$ and $128$ the measured values are averaged over $2000$ configurations. The intersection point indicates that the critical point is at $J_c = 0.33145(5)$.}
	\label{fig:g05d20vm}
\end{figure}

\begin{figure}[tbp]
	\begin{center}
		\includegraphics[width=\linewidth]{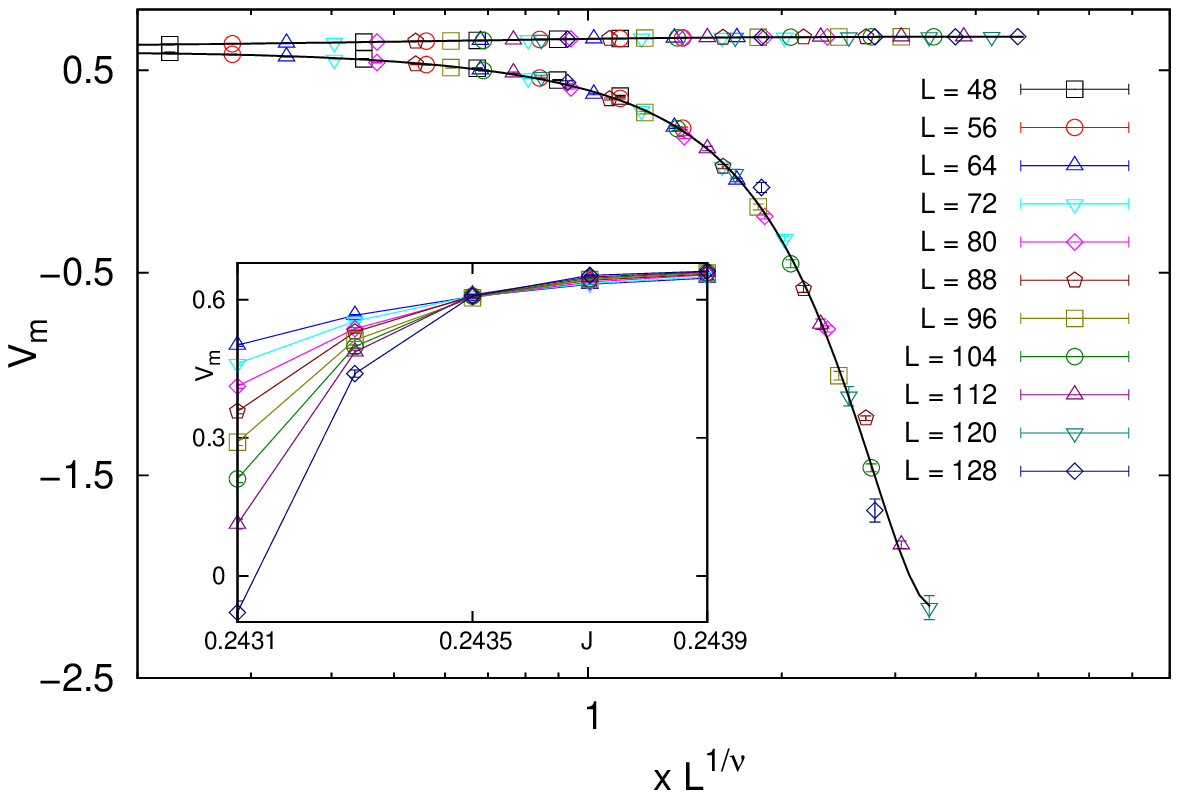}
	\end{center}
	\caption{(color online). Data collapse of the magnetic cumulant for the parameter set $(g,\Delta) = (0.1,0.1)$ with $N = 3, J_c = 0.24350$ and $\nu = 0.69$ for $L = 48, 56, \dots, 128$. The upper (lower) branch corresponds to the ordered (disordered) phase. The inset shows the magnetic cumulant $V_m$ versus $J$. The measured values for system sizes of up to $L = 112$ are averaged over $5000$ configurations. For $L = 120$ and $128$ the measured values are averaged over $2000$ configurations. The intersection point indicates that the critical point is at $J_c = 0.24350(5)$.}
	\label{fig:g10d10vm}
\end{figure}

A useful quantity that we calculate is the lowest-order magnetic cumulant,~\cite{binderPRB1984,challaPRB1986}
\begin{equation}\label{eq:magnetic_cumulant}
	V_m = 1 - \frac{[\langle m^4 \rangle]}{3[\langle m^2 \rangle]^2},
\end{equation}
where $m$ is the magnetization of the system given by
\begin{equation}
	m = \frac{1}{N}\left[\langle \sum_{\alpha=1}^N |m_\alpha| \rangle\right].
\end{equation}
We use the symmetry between spins of different colors to increase accuracy. In the disordered phase, $J < J_c$, one can show that~\cite{binderZPB1981,binderPRL1981} $V_m \propto L^{-D} \to 0$ as $L \to \infty$. In the ordered phase, $J > J_c$, we have spontaneous magnetization centered at $\pm m$, and therefore $V_m \to 2/3$. At the critical point, $J = J_c$, the magnetic cumulant approaches a fixed point.~\cite{binderZPB1981,binderPRL1981} Hence, we can extract the value of $J_c$ without any estimate of the critical exponents.
The insets in Figs.~\ref{fig:g10d20vm}, \ref{fig:g05d20vm}, and \ref{fig:g10d10vm} show the calculated magnetic cumulants, Eq. \eqref{eq:magnetic_cumulant}, for the parameter sets $(g,\Delta) \in \{(0.1,0.2), (0.05,0.2), (0.1,0.1)\}$.

According to the finite-size scaling, the singular part of energy density of the infinite-size lattice near the critical point is
\begin{eqnarray}\label{eq:scaling_function}
	f(x,H;L) = L^{-D}\mathcal F(axL^{1/\nu},bHL^{\delta/\nu})+\dots,
\end{eqnarray}
where $a$ and $b$ are metric factors making the scaling function $\mathcal F$ universal. $x$ is the reduced coupling constant given by $|J-J_c|/J_c$, and $\nu$ \textit{and} $\delta$ are the static critical exponents. From Eq.~\eqref{eq:scaling_function}, we can derive the scaling form of various thermodynamic quantities by taking appropriate derivatives. For the magnetic cumulant, we get
\begin{equation}\label{eq:magnetic_cumulant_scaled}
	V_m = \mathcal V (xL^{1/\nu}).
\end{equation}
At the critical point, the quantity
\begin{equation}
	\left.x\right|_{J=J_c} \equiv \left.\frac{|J-J_c|}{J_c}\right|_{J=J_c} = 0.
\end{equation}
Therefore, the argument of the function $\mathcal V$ in Eq.~\eqref{eq:magnetic_cumulant_scaled} vanishes. Hence, all curves cross at a single point as shown in the insets of Figs.~\ref{fig:g10d20vm}, \ref{fig:g05d20vm}, and \ref{fig:g10d10vm}.

After we have found $J_c$, the correlation length critical exponent, $\nu$, can straightforwardly be found using the scaled magnetic cumulant as given in Eq. \eqref{eq:magnetic_cumulant_scaled}. We fix $J_c$ within the statistical error bar and vary $\nu$. For each value of $\nu$, we fit the collapsed data to a polynomial and look for the least value of $\chi^2$. For lattice sizes $L = 48, 56, \dots, 128$ and the parameter set $(g, \Delta) = (0.1, 0.2)$, we found $J_c = 0.23287(3)$ and we obtained the best data collapse for the critical exponent $\nu = 0.81^{+0.02}_{-0.02}$ as shown in Fig.~\ref{fig:g10d20vm}. Similarly, in Fig.~\ref{fig:g05d20vm}, we show the data collapse corresponding to the parameter set $(g, \Delta) = (0.05, 0.20)$ with $J_c = 0.33145(5)$ and $\nu = 0.82^{+0.04}_{-0.02}$. For the parameter set $(g, \Delta) = (0.1, 0.1)$, we find $J_c = 0.24350(5)$ and $\nu = 0.69^{+0.01}_{-0.00}$ as depicted in Fig.~\ref{fig:g10d10vm}.


\begin{figure}[tbp]
	\begin{center}
		\includegraphics[width=\linewidth]{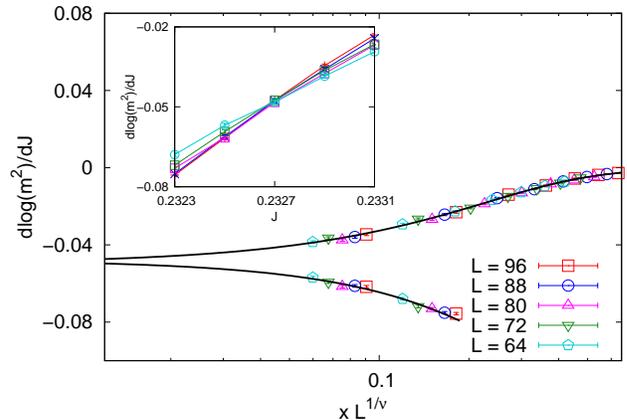}
	\end{center}
	\caption{(color online). Data collapse of the logarithmic derivative of the magnetization for the parameter set $(g,\Delta) = (0.1,0.2)$ with $N=3$, $J_c = 0.23270$ and $\nu = 0.98$ for $L = 96, 88, \dots, 64$. The upper (lower) branch corresponds to the ordered (disordered) phase. The inset shows the logarithmic derivative of the magnetization $d\log(m^2)/dJ$ versus $J$. The measured values are averaged over $5000$ configurations. The intersection point is at $J_c = 0.23270(10)$.}
	\label{fig:g10d20lm}
\end{figure}

\begin{figure}[tbp]
	\begin{center}
		\includegraphics[width=\linewidth]{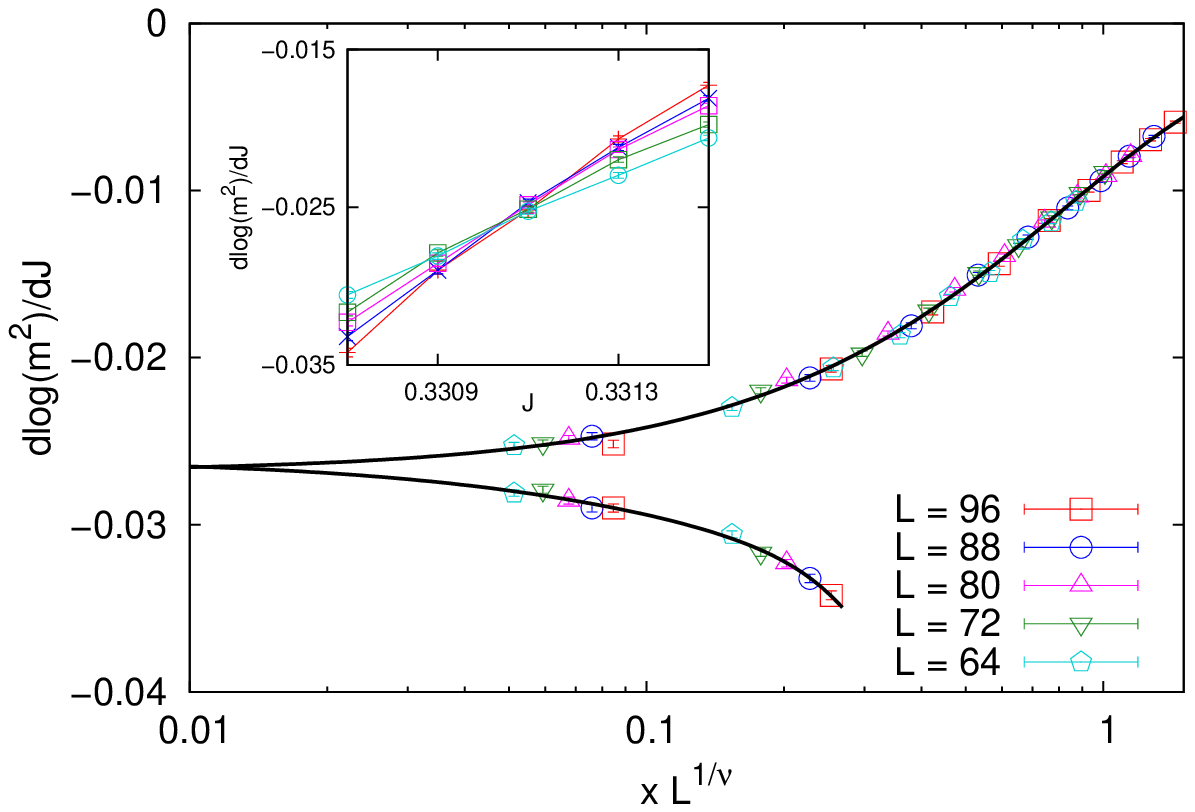}
	\end{center}
	\caption{(color online). Data collapse of the logarithmic derivative of the magnetization for the parameter set $(g,\Delta) = (0.05,0.2)$ with $N=3$, $J_c = 0.33100$ and $\nu = 0.81$ for $L = 64, 72, \dots, 96$. The upper (lower) branch corresponds to the ordered (disordered) phase. The inset shows the logarithmic derivative of the magnetization $d\log(m^2)/dJ$ versus $J$. The measured values are averaged over $5000$ configurations. The intersection point is at $J_c = 0.33100(5)$}
	\label{fig:g05d20lm}
\end{figure}

\begin{figure}[tbp]
	\begin{center}
		\includegraphics[width=\linewidth]{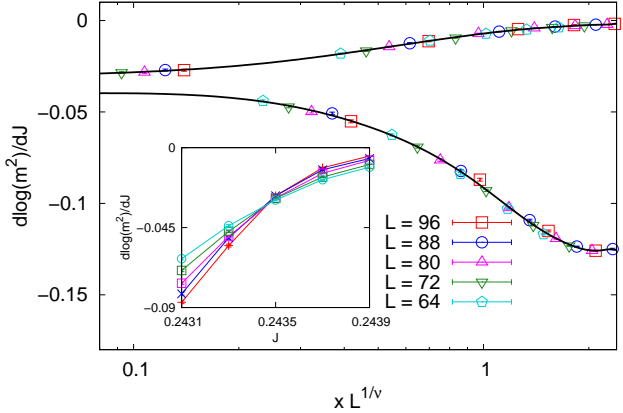}
	\end{center}
	\caption{(color online). Data collapse of the logarithmic derivative of the magnetization for the parameter set $(g,\Delta) = (0.1,0.1)$ with $N=3$, $J_c = 0.24345$ and $\nu = 0.70$ for $L = 64, 72, \dots, 96$. The upper (lower) branch corresponds to the ordered (disordered) phase. The inset shows the logarithmic derivative of the magnetization $d\log(m^2)/dJ$ versus $J$. The measured values are averaged over $5000$ configurations. The intersection point is at $J_c = 0.24345(5)$.}
	\label{fig:g10d10lm}
\end{figure}

Another thermodynamic quantity that we found useful in determining the critical point and critical exponent $\nu$ is the logarithmic derivative of the $n$-th power of magnetization~\cite{ferrenbergPRB1991}
\begin{equation}\label{eq:logarithmic_derivative}
	\frac{\partial}{\partial J}\log\langle m^n \rangle = \frac{1}{m^n} \frac{\partial}{\partial J}\langle m^n\rangle = \frac{\langle m^n E \rangle}{\langle m^n \rangle} - \langle E \rangle.
\end{equation}
This quantity has similar finite-size scaling as the magnetic cumulant.~\cite{ferrenbergPRB1991} We only calculate the logarithmic derivative of the magnetization squared for system sizes of $L = 64, 72, \dots, 96$. We use this to find a second estimate for the critical point, $J_c$, and the correlation length critical exponent, $\nu$. As before, we calculate this quantity for the three parameter sets $(g,\Delta) \in \{(0.1,0.2), (0.05, 0.20), (0.1,0.1)\}$. We extract the critical point by finding the crossing point of $d\log(m^2)/dJ$ as a function of $J$ for different lattice sizes as shown in the insets of Figs.~\ref{fig:g10d20lm}, \ref{fig:g05d20lm}, and \ref{fig:g10d10lm}. For the parameter set $(g, \Delta) = (0.1, 0.2)$, we find $J_c = 0.23270(10)$. For $(g, \Delta) = (0.05, 0.2)$, we find $J_c = 0.33100(5)$. Finally, for $(g, \Delta) = (0.1, 0.1)$, we find $J_c = 0.24345(5)$.

We extract the critical exponent $\nu$ from the logarithmic derivative of the magnetization squared data using analogous procedure as explained before after Eq.~\eqref{eq:magnetic_cumulant_scaled}. The data collapse of the logarithmic derivative of the magnetization squared is shown in Figs.~\ref{fig:g10d20lm}--\ref{fig:g10d10lm}. The obtained critical exponents for different values of $g$ and $\Delta$ are summarized in Table~\ref{tbl:exponents}.


\begin{figure}[tbp]
	\begin{center}
		\includegraphics[width=\linewidth]{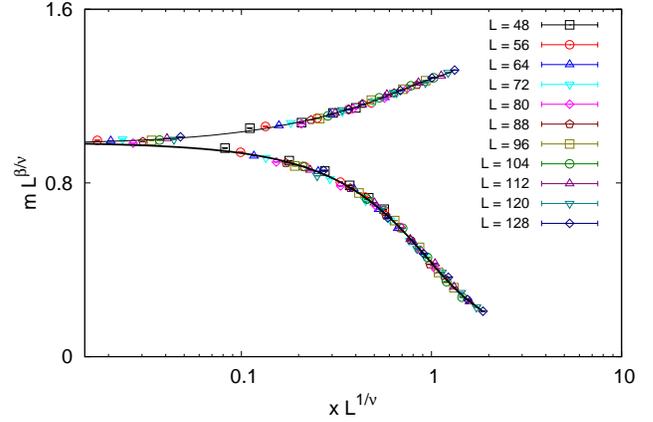}
		\includegraphics[width=\linewidth]{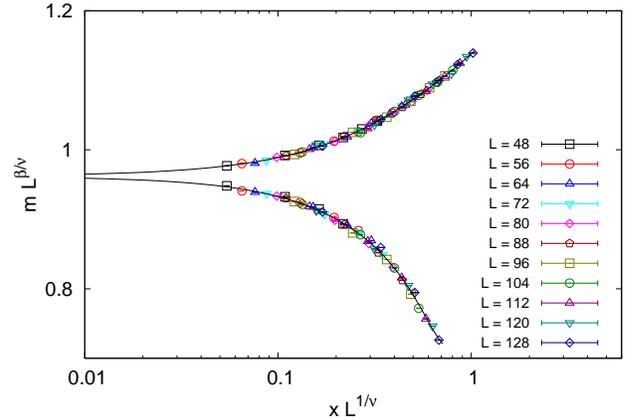}
		\includegraphics[width=\linewidth]{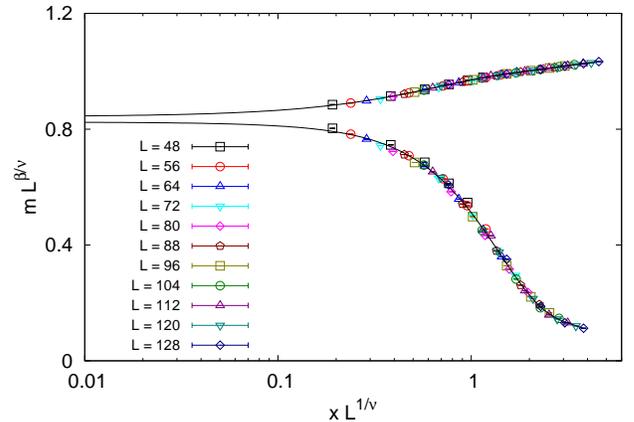}
	\end{center}
	\caption{(color online). Data collapse of magnetization for system sizes $L = 48, 56, \dots, 128$. The top figure corresponds to the parameter set $(g,\Delta) = (0.1,0.2)$ with $N=3, J_c = 0.23287$, and $\beta/\nu \approx 0.12$. The middle figure corresponds to the parameter set $(g,\Delta) = (0.05,0.2)$ with $N=3, J_c = 0.33145$, and $\beta/\nu \approx 0.10$. The bottom figure corresponds to the parameter set $(g,\Delta) = (0.1,0.1)$ with $N=3, J_c = 0.24350$, and $\beta/\nu \approx 0.05$. The magnetization values for system sizes of up to $L = 112$ are averaged over $5000$ configurations. For the system sizes $L = 120$ and $128$, the magneziation values are averaged over only $2000$ configurations. The upper (lower) branch in each figure corresponds to the ordered (disordered) phase.}
	\label{fig:mfss}
\end{figure}

We find the magnetic critical exponent, $\beta$, using the scaled magnetization function
\begin{equation}\label{eq:magnetization_scaled}
	m = L^{-\beta/\nu} \mathcal M(xL^{1/\nu}).
\end{equation}
Our analysis to obtain $\beta$ is similar to our analysis for finding $\nu$ with the exception that this time we fix $J_c$ within the statistical error bar and vary $\nu$ and $\beta$. We fit the collapsed data to a polynomial and find the least $\chi^2$ value. The data collapse of magnetization for different values of $g$ and $\Delta$ is shown in Fig.~\ref{fig:mfss} and the obtained critical exponents $\nu$ and $\beta$ are summarized in Table~\ref{tbl:exponents}.

\section{Conclusion}\label{sec:conclusion}
In this work, we performed an extensive MC calculation on the quenched bond-disordered three-color Ashkin-Teller model with large lattice sizes, up to $128 \times 128$. The calculation supports our earlier results for smaller system sizes,~\cite{bellafardPRL2012} namely, (1) the quenched disorder rounds a first-order phase transition to a critical point, (2) the critical exponents are not in the Ising universality class ($\nu=1$ and $\beta=1/8$), and (3) the critical exponents depend on the disorder and the four-spin coupling strengths. In our present work, we have found the critical exponent of the correlation length, $\nu$, through finite-size scaling of two different quantities: (1) the magnetic cumulant and (2) the logarithmic derivative of the magnetization squared. The obtained values are within the error bars. Table~\ref{tbl:exponents} summarizes our computed exponents.

\begin{table}[tbp]
\renewcommand{\arraystretch}{1.5}
\footnotesize
\centering
\begin{tabularx}{\columnwidth}{@{\extracolsep{\stretch{1}}}l |l |l |l@{}}
\hline\hline
$(g,\Delta)$	&	$(0.10, 0.20)$	&	$(0.05, 0.20)$	&	$(0.10, 0.10)$\\
\hline\hline
$V_m$			&	$\nu = 0.81^{+0.02}_{-0.02}$ &	$\nu = 0.82^{+0.04}_{-0.02}$	&	$\nu = 0.69^{+0.01}_{-0.00}$\\
\hline
$d\log m^2/dJ$	&	$\nu = 0.98^{+0.17}_{-0.07}$&	$\nu = 0.81^{+0.04}_{-0.04}$	&	$\nu = 0.70^{+0.02}_{-0.02}$\\
\hline
$m$				&	$\nu = 0.82^{+0.02}_{-0.02}$	&	$\nu = 0.86^{+0.03}_{-0.03}$	&	$\nu = 0.71^{+0.00}_{-0.09}$\\
				&	$\beta =  0.10^{+0.01}_{-0.01}$	&	$\beta =  0.09^{+0.00}_{-0.01}$	&	$\beta = 0.04^{+0.01}_{-0.01}$\\
\hline\hline
\end{tabularx}
\setlength{\abovecaptionskip}{5pt}
\caption{The values of the critical exponents extracted by finite-size scaling of magnetic cumulant, $V_m$, logarithmic derivative of the magnetization squared, $d\log m^2/dJ$, and magnetization, $m$, for different values of four-spin coupling constant, $g$, and disorder, $\Delta$. These critical exponents are to be contrasted with the Ising exponents for which $\nu = 1$ and $\beta = 1/8$. The results for $d\log m^2/dJ$ are for system sizes $96\times 96$ and the results for $V_m$ and $m$ are for system sizes $128\times 128$.}
\label{tbl:exponents}
\end{table}

\begin{figure}[tbp]
	\begin{center}
		\includegraphics[width=\linewidth]{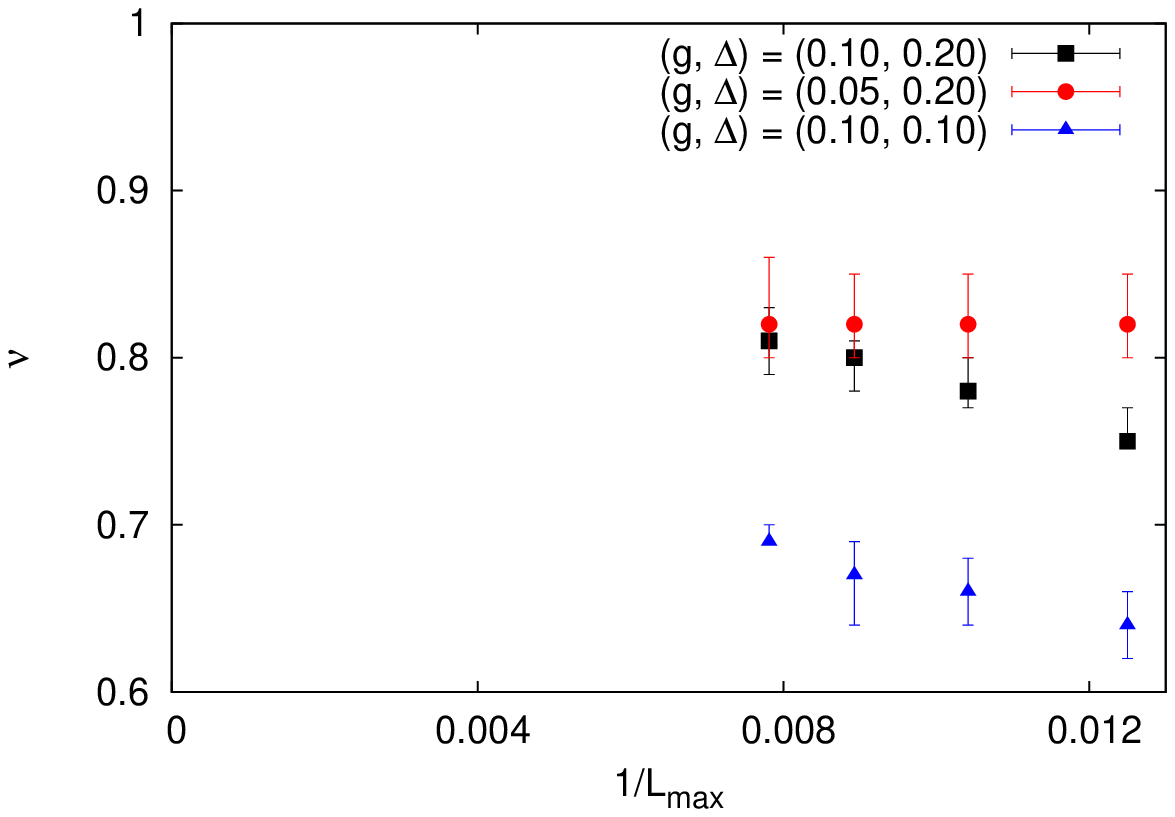}
	\end{center}
	\caption{(color online). Trend of the critical exponent $\nu$ as a function of $1/L_{\rm max}$. $L_{\rm max}$ is the largest of the examined system sizes. The exponents $\nu$ are extracted by finite-size scaling of the \emph{magnetic cumulant}. The explicit values of $\nu$ are reported in Table~\ref{tbl:nutrend}. For the system sizes of up to $112\times 112$, the magnetic cumulants are averaged over $5000$ disorder configurations. For the largest system size, $128\times 128$, the magnetic cumulants are averaged over $2000$ disorder configurations.}
	\label{fig:nutrend}
\end{figure}

\begin{figure}[tbp]
	\begin{center}
		\includegraphics[width=\linewidth]{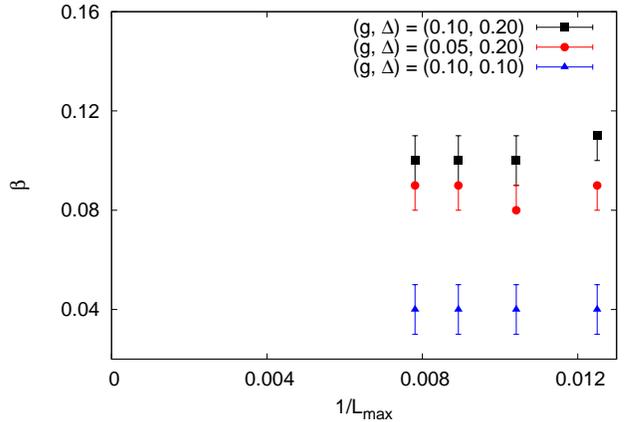}
	\end{center}
	\caption{(color online). Trend of the critical exponent $\beta$ as a function of $1/L_{\rm max}$. $L_{\rm max}$ is the largest of the examined system sizes. The exponents $\beta$ are extracted by finite-size scaling of the magnetization. The explicit values of $\beta$ are reported in Table~\ref{tbl:betatrend}. For the system sizes of up to $112\times 112$, the magnetic cumulants are averaged over $5000$ disorder configurations. For the largest system size, $128\times 128$, the magnetic cumulants are averaged over $2000$ disorder configurations.}
	\label{fig:betatrend}
\end{figure}

A remarkable point to notice is that the correlation length critical exponent, $\nu$, and the magnetization critical exponent, $\beta$, change as the disorder strength or the four-spin coupling constant varies. 
To examine the finite-size dependence of the critical exponents, we extract the values of the $\nu$ and $\beta$ for different values of the system sizes. In Tables~\ref{tbl:nutrend} and~\ref{tbl:betatrend}, we report the values of $\nu$ and $\beta$ obtained for various system sizes. In Figs.~\ref{fig:nutrend} and~\ref{fig:betatrend} we plot the values of $\nu$ and $\beta$ as a function of $L_{\rm max}^{-1}$. The data points seem to lie well on a straight line. 
For the $(g,\Delta) = (0.10,0.20)$ case, one can argue that in the thermodynamic limit, $L_{\rm max} \to \infty$, the exponent $\nu \to 1$ (Ising), however, the exponent $\beta$ for this case does not seem to approach $1/8$ for it to fall within the Ising universality class. The exponents of the other two parameter sets, $(g,\Delta) = (0.05,0.20)$ and $(0.10,0.10)$, also appear to differ from the Ising exponents even when we extrapolate to $L_{\rm max} \to \infty$.

\begin{table}[tbp]
\renewcommand{\arraystretch}{1.5}
\footnotesize
\centering
\begin{tabularx}{\columnwidth}{@{\extracolsep{\stretch{1}}}l |l |l |l@{}}
\hline\hline
$(g,\Delta)$		&	$(0.10, 0.20)$		&	$(0.05, 0.20)$	&	$(0.10, 0.10)$\\
\hline\hline
$L = 48, 56, \dots, 80$		&	$\nu = 0.75^{+0.02}_{-0.00}$	&	$\nu = 0.82^{+0.03}_{-0.02}$ &	$\nu = 0.64^{+0.02}_{-0.02}$\\
\hline
$L = 48, 56, \dots, 96$		&	$\nu = 0.78^{+0.02}_{-0.01}$	&	$\nu = 0.82^{+0.03}_{-0.02}$ &	$\nu = 0.66^{+0.02}_{-0.02}$\\
\hline
$L = 48, 56, \dots, 112$	&	$\nu = 0.80^{+0.01}_{-0.02}$	&	$\nu = 0.82^{+0.03}_{-0.02}$ &	$\nu = 0.67^{+0.02}_{-0.03}$\\
\hline
$L = 48, 56, \dots, 128$	&	$\nu = 0.81^{+0.02}_{-0.02}$	&	$\nu = 0.82^{+0.04}_{-0.02}$ &	$\nu = 0.69^{+0.01}_{-0.00}$\\
\hline\hline
\end{tabularx}
\setlength{\abovecaptionskip}{5pt}
\caption{The size dependence of the critical exponent $\nu$ for different values of four-spin coupling constants, $g$, and disorder, $\Delta$. The values of $\nu$ are extracted by finite-size scaling of the \emph{magnetic cumulant} for different system sizes. For the system sizes of up to $112\times 112$, the magnetic cumulants are averaged over $5000$ disorder configurations. For the system sizes $120\times 120$ and $128\times 128$, the magnetic cumulants are averaged over $2000$ disorder configurations.}
\label{tbl:nutrend}
\end{table}

\begin{table}[tbp]
\renewcommand{\arraystretch}{1.5}
\footnotesize
\centering
\begin{tabularx}{\columnwidth}{@{\extracolsep{\stretch{1}}}l |l |l |l@{}}
\hline\hline
$(g,\Delta)$		&	$(0.10, 0.20)$		&	$(0.05, 0.20)$	&	$(0.10, 0.10)$\\
\hline\hline
$L = 48, 56, \dots, 80$		&	$\beta =  0.11^{+0.00}_{-0.01}$	&	$\beta =  0.09^{+0.00}_{-0.01}$ &	$\beta =  0.04^{+0.01}_{-0.01}$\\
\hline
$L = 48, 56, \dots, 96$		&	$\beta =  0.10^{+0.01}_{-0.01}$	&	$\beta =  0.08^{+0.01}_{-0.00}$ &	$\beta =  0.04^{+0.01}_{-0.01}$\\
\hline
$L = 48, 56, \dots, 112$	&	$\beta =  0.10^{+0.01}_{-0.01}$	&	$\beta =  0.09^{+0.00}_{-0.01}$	&	$\beta =  0.04^{+0.01}_{-0.01}$\\
\hline
$L = 48, 56, \dots, 128$	&	$\beta =  0.10^{+0.01}_{-0.01}$	&	$\beta =  0.09^{+0.00}_{-0.01}$	&	$\beta =  0.04^{+0.01}_{-0.01}$\\
\hline\hline
\end{tabularx}
\setlength{\abovecaptionskip}{5pt}
\caption{The size dependence of the critical exponent $\beta$ for different values of four-spin coupling constants, $g$, and disorder, $\Delta$. The values of $\beta$ are extracted by finite-size scaling of the magnetization for different system sizes. For the system sizes of up to $112\times 112$, the magnetic cumulants are averaged over $5000$ disorder configurations. For the system sizes $120\times 120$ and $128\times 128$, the magnetic cumulants are averaged over $2000$ disorder configurations.}
\label{tbl:betatrend}
\end{table}

We believe that  the quenched bond-disordered three-color Ashkin-Teller model does not belong to the Ising universality class. In addition, a point that merits further discussion is that the lower bound on the correlation length critical exponent derived by \citeauthor{chayesPRL1986},~\cite{chayesPRL1986} $2/D \le \nu$, appears to be violated. Those transitions that are first-order in the pure system but are rendered continuous by addition of bond disorder may deserve further attention.~\cite{chayesPrivateCommunication}

\section{Acknowledgment}\label{sec:acknowledgment}
All computations were carried out  at UCLA on Hoffman2 and ETHZ Brutus high-performance computing clusters. A.B. and S.C. thank the National Science Foundation, Grant No. DMR-1004520 for support. S.C. also acknowledges support from the funds from David S. Saxon Presidential Chair at UCLA. H.G.K. acknowledges support from the National Science Foundation, Grant No. DMR-1151387. A.B. thanks Dr. Ruben Andrist for helpful discussions. M.T. and S.C.  thank the  Aspen center for physics for its hospitality and support through a grant NSF-PHY-1066293.

\end{document}